\documentclass[10 pt, letter]{article}
\usepackage{graphicx}
\usepackage{caption}
\usepackage{amsmath}
\usepackage{amsthm}
\usepackage{enumerate}
\usepackage{subfig}
\usepackage{float}
\usepackage{amsfonts}
\usepackage{amssymb}
\usepackage{anysize}
\usepackage{helvet}
\usepackage{setspace}
\usepackage{multirow}
\usepackage{tabularx}
\usepackage{tabulary}
\usepackage{longtable}
\usepackage{float}
\usepackage{mathtools}
\title{\bf Generalization of the Force Approach to Radiation Reaction}
\author{Gustavo V. L\'opez\footnote{gulopez@udgserv.cencar.udg.mx}~\\
 Departamento de F\'{i}sica, Universidad de Guadalajara,\\
 Blvd. Marcelino Garc\'{i}a Barragan y Calzada Ol\'{i}mpica, 44200 Guadalajara, Jalisco, Mexico}
\begin{document}
\maketitle

%\centerline{\large\bf Abstract}
\begin{abstract}
\noindent
A generalization of the force approach to radiation reaction is given, taken into consideration an arbitrary motion of the charged particle . The expression obtained brings about the expression already given for the linear an the circular acceleration cases.
\end{abstract}
\newpage
\section{ Introduction}
Recently, it appeared an alternative approach [1] to the old standing problem of radiation reaction [2,3,4,5]. This new approach is based on expressing the radiation reaction force in terms of the external force acting on the charged particle, which it is the responsible for its acceleration. Although, this approach seems to point in the right direction, experimental verification  of this theoretical idea is required. In reference [1], the approach is shown for linear and circular acceleration of a charged particle. In this paper, a generalization of the idea is carried out, using the same method presented in that paper. An expression is shown that represent this generalization, and for the linear an circular acceleration cases is reduced to the expression found in reference [1]. 
\section{Acceleration and force relation}
The relativistic equation to describe the motion of a charge particle under an external force ${\bf F}$ is given by [6]  
\begin{equation}\label{eq1}
\frac{d\gamma m{\bf v}}{dt}={\bf F},
\end{equation}
where $m$ and ${\bf v}$ are the mass and the velocity of the charged particle, and $\gamma$ is the usual relativistic function
\begin{equation}
\gamma=(1-\beta^2)^{-1/2},
\end{equation}
being $\beta$ the normalized speed of the charged particle with respect the speed of light "c",
\begin{equation}
\beta=\frac{v}{c}=\frac{1}{c}\sqrt{v_x^2+v_y^2+v_z^2}.
\end{equation}
Making the differentiation in (\ref{eq1}), this equation can be written as
\begin{equation}
m\gamma^3\begin{pmatrix}1-\beta_y^2-\beta_z^2&\beta_x\beta_y&\beta_x\beta_z\\Ê\\
\beta_y\beta_x&1-\beta_x^2-\beta_z^2&\beta_y\beta_z\\Ê\\
\beta_z\beta_x&\beta_y\beta_y&1-\beta_x^2-\beta_y^2\end{pmatrix}\begin{pmatrix}\dot v_x\\Ê\\ \dot v_y\\Ê\\ \dot v_x\end{pmatrix}={\bf F},
\end{equation}
where $\beta_i=v_i/c$ i=x,y,z. Inverting this matrix and making some rearrangements, it follows that
\begin{equation}\label{vF}
\dot{\vec\beta}=\frac{1}{mc\gamma}\aleph{\bf F},
\end{equation}
where $\aleph$ is the matrix 
\begin{equation}
\aleph=\begin{pmatrix}1-\beta_x^2&-\beta_x\beta_y&-\beta_x\beta_z\\Ê\\ -\beta_y\beta_x&1-\beta_y^2&-\beta_y\beta_z\\Ê\\
-\beta_z\beta_x&- \beta_z\beta_y& 1-\beta_z^2\end{pmatrix}.
\end{equation}
This is the expression with makes the relation between the normalized acceleration with the external force.
\section{ Radiation Reaction Force}
It is well known that the power radiated per solid angle of an accelerated charged particle  of charge "q" is [7]
\begin{equation}
\frac{dP}{d\Omega}=\frac{q^2}{4\pi c}\frac{\bigr|\hat{\bf R}\times[(\hat{\bf R}-\vec\beta)\times\dot{\vec\beta}]\bigr|^2}{(1-\hat{\bf R}
\cdot\vec\beta)^5},
\end{equation}
where $\hat{\bf R}$ is the unitary vector that goes from the position of the charge at the retarded time ($t'=t+R/c$) to the observer position. So, using (\ref{vF})  on this expression, it follows that
\begin{equation}
\frac{dP}{d\Omega}=\frac{q^2}{4\pi m^2c^3\gamma^2}\frac{\bigr|\hat{\bf R}\times[(\hat{\bf R}-\vec\beta)\times\aleph{\bf F}]\bigr|^2}{(1-\hat{\bf R}\cdot\vec\beta)^5},
\end{equation}
Integrating with respect the solid angle and with respect the time in the intervale $[0,t]$, the energy radiated by the charged particle during this time is
\begin{equation}
U(t)=\frac{q^2}{4\pi m^2c^3}\int_0^t\frac{dt}{\gamma^2}
\int_{\Omega}\frac{\bigr|\hat{\bf R}\times[(\hat{\bf R}-\vec\beta)\times\aleph{\bf F}]\bigr|^2}{(1-\hat{\bf R}\cdot\vec\beta)^5}d\Omega.
\end{equation}
which can be written in terms of angles as
\begin{equation}\label{urra}
U(t)=\frac{q^2}{4\pi m^2c^3}\int_0^t\frac{|\aleph{\bf F}|^2dt}{\gamma^2}
\int_{\Omega}\frac{\bigr|{\bf e}_1\sin\theta_1-{\bf e}_2\beta\sin\theta_2\bigr|^2}
{(1-\beta\cos\theta)^5}d\Omega,
\end{equation}
where ${\bf e}_1$ is the unitary vector in the direction $\hat{\bf R}\times(\hat{\bf R}\times\aleph{\bf F})$ and ${\bf e}_2$ is the unitary vector in the direction $\hat{\bf R}\times(\vec\beta\times\aleph{\bf F})$, $\theta_1$ is the angle between the vectors $\hat{\bf R}$ and $\hat{\bf R}\times\aleph{\bf F}$, and $\theta_2$ is the angle between the vectors  $\hat{\bf R}$ and $\vec\beta\times\aleph{\bf F}$. One can choose our reference system such that $\theta$ be the angle related with the solid angle coordinates. The angles $\theta_1$ and $\theta_2$ depends on the solid angle coordinates ($\theta, \phi$). \\Ê \\
Assume now that this energy lost is due to the work done by a nonconservative radiation reaction force, to move the charged particle from the position ${\bf x}_0$ a the time $t=0$, to the position ${\bf x}$ at the time t. So, one would have
\begin{equation}
U(t)=\int_{\bf x_0}^{\bf x}{\bf F}_{rad}\cdot d{\bf x}.
\end{equation} 
However, one has that $d{\bf x}={\bf v} dt$. Then, it follows that
\begin{equation}\label{urad}
U(t)=\int_0^t{\bf F}_{rad}\cdot{\bf v} dt.
\end{equation}
Equaling (\ref{urra}) and (\ref{urad}), and since the resulting expression is valid for any time intervale $[0,t]$ on the real line, one obtains
\begin{equation}\label{rada}
F_{rad}=\frac{q^2|\aleph{\bf F}|^2}{4\pi m^2c^3\gamma^2v\cos\varphi}
\int_{\Omega}\frac{\bigr|{\bf e}_1\sin\theta_1-{\bf e}_2\beta\sin\theta_2\bigr|^2}
{(1-\beta\cos\theta)^5}d\Omega,
\end{equation}
where $v$ is the charged particle speed and $\varphi$ is the angle between the vectors ${\bf F}_{rad}$ and ${\bf v}$. Since ${\bf F}_{rad}$ must represent a force causing damping on the motion of the charged particle, ${\bf F}_{rad}$ must be point on the $-\hat{\bf n}$ direction, where $\hat{\bf n}={\bf v}/v$, meaning the the angle between ${\bf F}_{rad}$ and ${\bf v}$ must be $\varphi=\pi$. Therefore, one gets the following expression for radiation reaction force
\begin{equation}\label{final}
{\bf F}_{rad}=-\frac{q^2|\aleph{\bf F}|^2{\bf v}}{4\pi m^2c^3\gamma^2v^2}
\int_{\Omega}\frac{\bigr|{\bf e}_1\sin\theta_1-{\bf e}_2\beta\sin\theta_2\bigr|^2}
{(1-\beta\cos\theta)^5}d\Omega,
\end{equation}
In this way, the modified relativistic equation of motion of a charged particle under an arbitrary external force ${\bf F}$ is
\begin{equation}
\frac{d\gamma m{\bf v}}{dt}={\bf F}+{\bf F}_{rad},
\end{equation}
or
\begin{equation}\label{modeq}
\frac{d\gamma m{\bf v}}{dt}={\bf F}-\frac{q^2|\aleph{\bf F}|^2{\bf v}}{4\pi m^2c^3\gamma^2v^2}
\int_{\Omega}\frac{\bigr|{\bf e}_1\sin\theta_1-{\bf e}_2\beta\sin\theta_2\bigr|^2}
{(1-\beta\cos\theta)^5}d\Omega.
\end{equation}
One must  point out that if the external force ${\bf F}$ is zero, the radiation reaction force ${\bf F}_{rad}$ is also zero, and noncausal preacceleration exists since the charged particle will have constant velocity, as the experiments indicate so far.  
\section{Special Cases}
As one can immediately see from  (\ref{final}), the integration over the solid angle is not in general a trivial matter, and numerical  integration may be required. However, there are two cases where this integration can be done without any difficulty, and these cases are presented below.\\Ê\\
a) Linear acceleration case: In this case one has that $\aleph {\bf F}$ is parallel to $\vec\beta$ ( ${\bf e}_2={\bf 0}$)  and there is not dependence on $\phi$, this integration is well known [7], and choosing $\vec\beta=(0,0,\beta_z)$ and ${\bf F}=(0,0,F)$, the resulting equation is
\begin{equation}
\frac{d\gamma mv}{dt}=F-\frac{\lambda_0F^2}{v},
\end{equation}   
where $\gamma=(1-\beta_z^2)^{-1/2}$, and $\lambda_0$ is defined as
\begin{equation}
\lambda_0=\frac{2q^2}{3m^2c^3}.
\end{equation}
This equation is the same as that proposed for this case on reference [1].\\Ê\\
b) Circular acceleration case: In this case one has that $\aleph{\bf F}$ and $\vec\beta$ are orthogonal. The integration is also well known [7], and choosing $\vec\beta=(\beta_x,\beta_y,0)$ and ${\bf F}=(F_x,F_y,0)$, one arrives to the following equation
\begin{equation}
\frac{d\gamma m{\bf v}}{dt}={\bf F}-\frac{\lambda_0 F^2}{v^2\gamma^2}{\bf v},
\end{equation}
where $\gamma=(1-\beta_x^2-\beta_y^2)^{-1/2}$, $v^2=v_x^2+v_y^2$, and the motion is reduced to the plane (x,y). This expression is the same as that one proposed on reference [1].
\section{ Results}
Following the same approach of reference [1], a generalization was given to the force approach to radiation reaction, expression (\ref{modeq}). This generalization has the same main property of the force approach,  wherever the external force is zero, the acceleration of the charged particle is zero, and the radiation reaction force is zero too. 
\end{document}